\documentclass[12pt,twocolumn]{emulateapj}

\def\alwaysmath#1{\ifmmode{#1}\else{$#1$}\fi}

\shortauthors{MEN\'{E}NDEZ-DELMESTRE ET AL. 2006}
\shorttitle{MID-IR SPECTROSCOPY OF SUBMILLIMETER GALAXIES}

\begin{document}

\title{Mid-Infrared Spectroscopy of High Redshift Submillimeter Galaxies: First Results} \author{\sc Kar\'{i}n
Men\'{e}ndez-Delmestre\altaffilmark{1}, Andrew W. Blain\altaffilmark{1}, Dave M. Alexander\altaffilmark{2}, Ian Smail\altaffilmark{2}, Lee Armus\altaffilmark{3}, Scott C. Chapman\altaffilmark{4,5},  D. T. Frayer\altaffilmark{3}, Rob J. Ivison\altaffilmark{6,7}, H. I. Teplitz\altaffilmark{3}}

\altaffiltext{1}{California Institute of Technology, MC 105-24, Pasadena, CA 91125}

\altaffiltext{2}{Institute for Computational Cosmology, Durham University, Durham DH1 3LE, UK}

\altaffiltext{3}{Spitzer Science Center, MC 220-6, California Institute of Technology, Pasadena, CA 91125}

\altaffiltext{4}{Institute of Astronomy, Madingley Road, Cambridge, CB3\,0HA, U.K.}

\altaffiltext{5}{CSA Fellow, U. of Victoria, Victoria BC, V8P 1A1 Canada}

\altaffiltext{6}{UK Astronomy Technology Centre, Blackford Hill, Edinburgh EH9 3HJ}

\altaffiltext{7}{Institute for Astronomy, Blackford Hill, Edinburgh EH9 3HJ}

\email{km@astro.caltech.edu}

\begin{abstract}
We present mid-infrared spectra of 5 submillimeter galaxies at $z=0.65-2.38$ taken with the  \textit{Spitzer Space Telescope}. Four of these sources, at $z\,\lesssim\,1.5$, have strong PAH features and their composite spectrum is well fitted by an M82-like spectrum with an additional power-law component consistent with that expected from AGN activity. Based on a comparison of the 7.7-$\mu$m PAH   equivalent width and the PAH-to-infrared luminosity ratio of these galaxies with local templates, we conclude that these galaxies host both star-formation and AGN activity, with star-formation dominating the bolometric luminosity. The source at $z\,=\,2.38$ displays a Mrk\,231-type broad feature at restframe $\sim$8\,$\mu$m that does not conform to the typical 7.7/8.6\,$\mu$m PAH complex in starburst galaxies, suggesting a more substantial AGN contribution. 
\end{abstract}


\section{INTRODUCTION}

Deep submillimeter-wave surveys  \citep{smail97, barger99, eales99, cowie02, scott02, borys03, webb03b} have uncovered a population of ultra-luminous infrared (IR) galaxies (ULIRGs;  $L_{\small\small\textrm{IR}} \textgreater\,2-5 \times 10^{12}\, $L$_{\odot}$) at z$\sim$2 \citep{blain02}. This observationally-defined population of submillimeter galaxies (SMGs) coincides with the epoch of peak global star formation and quasar activity, with a significant contribution to the global star formation rate density at $z=2-3$ not traced in the UV (\citealt{chapman05}, C05). Highly obscured by their dust content, the astrophysics of SMGs and the nature of their power source remain a challenge to address at optical and near-IR wavelengths (\citealt{chapman03}; C05; \citealt{swinbank04}). Deep X-ray studies suggest that $\sim 28-50\%$ of SMGs host an active galactic nucleus (AGN), although at face value it appears that the AGN does not dominate the bolometric luminosity and that powerful starbursts (SB) contribute more significantly to the total energy output (\citealt{alexander05}, A05). 

Less hindered by obscuration than shorter wavelengths, the mid-IR region boasts a number of spectral features, including: emission from Polycyclic Aromatic Hydrocarbons (PAHs) (e.g. restframe 6.2, 7.7, 8.6, 11.3 and 12.7\,$\mu$m), associated with star formation \citep{helou99} and typically absent in powerful AGN \citep{voit92}; silicate absorption at 9.7 and 18\,$\mu$m, which gives a measure of the obscuration by silicate dust grains along the line of sight to a small hot dust continuum source; and a hot dust continuum ($\lambda \lesssim 10\,\mu$m), likely to be dominated by an AGN. The strength of these features have been used in mid-IR surveys with the \textit{Infrared Space Observatory} \citep{genzel98, rigopoulou99, tran01, laurent00} to estimate the relative contributions of SB and AGN for the brightest local galaxies (e.g. \citealt{rigopoulou99}). The spectra of the high-redshift population remained unexplored in the mid-IR, until the advent of the \textit{Spitzer Space Telescope} and the unprecedented sensitivity of the Infrared Spectrograph (IRS; \citealt{houck04}). \citet{lutz05} (L05), \citet{yan05} (Y05), \citet{houck05}, \citet{desai06} and \citet{weedman06} have been among the pioneers in using IRS to study the mid-IR spectra of luminous sources at $z \sim1-3$, extending to higher redshifts the analysis that was previously only accessible for nearby galaxies.

We have an IRS program to study the range of mid-IR properties of a sample of 24 high-$z$ SMGs  with S$_{24 \mu\small\textrm{m}} \gtrsim 0.4$\,mJy, using the radio-identified sample with spectroscopic redshifts, compiled by C05. Here we present IRS spectra of the first five targets observed (SMM\,J221733 +001120, SMM\,J163659 +405728, SMM\,J030228 +000654, SMM\,J163639 +405636, and SMM\,J163650 +405735): four are at lower-$z$, with $z=0.65-1.5$, and one is at $z=2.38$. The low-$z$ targets cover wavelengths longwards of $10\,\mu$m and give insight into the longer mid-IR emission from SMGs; the full sample is more focused on $z \sim 2$ SMGs and hence probes shorter restframe wavelengths. This preliminary sample is otherwise representative of the SMG population, in terms of bolometric luminosity, dust temperature and submillimeter-to-radio flux ratio (C05).

\begin{figure*}\label{spectra}
\plotone{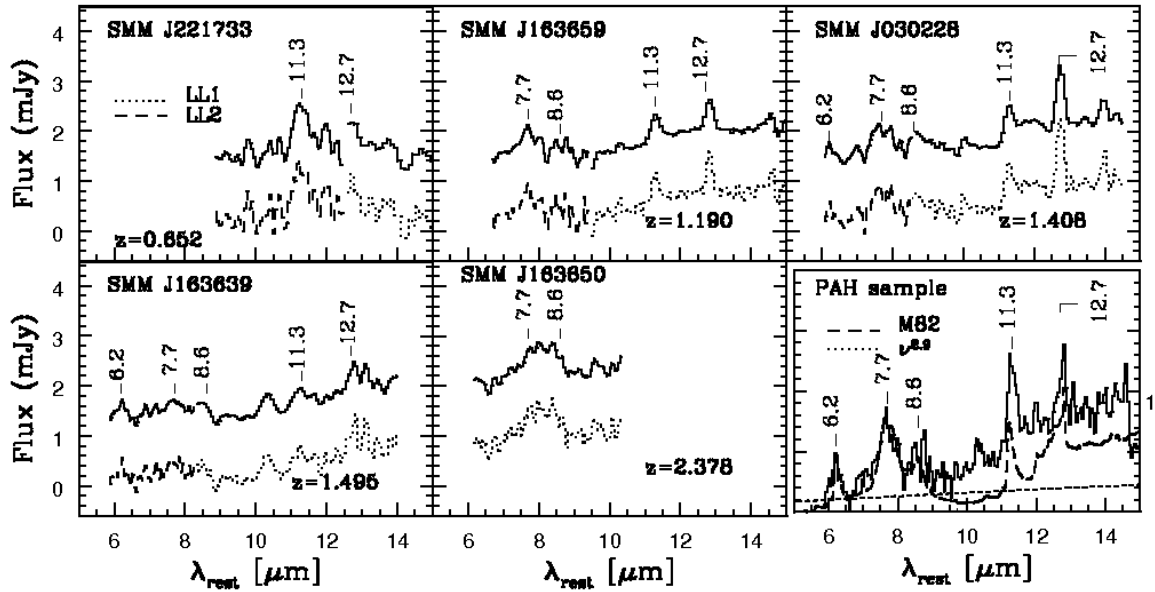}
\caption{1D $Spitzer$ IRS spectra for 5 SMGs and the composite spectrum of the PAH sample. For the 5 individual spectra, the lower curve represents the unsmoothed spectrum, with the first order (LL1: $\lambda_{obs} =19.5-38\,\mu$m) and second order (LL2: $\lambda_{obs} =14-21.3\,\mu$m) of the low-resolution mode shown in dotted and dashed lines, respectively. The upper curve shows the spectrum smoothed by 3 pixels and offset in flux for clarity. The various wavelengths of PAH emission features are indicated. We show the smoothed version of the composite spectrum for the PAH sample, together with the ISO SWS spectra of M82 (dashed line), smoothed to the resolution of IRS and normalized to the 7.7-$\mu$m PAH feature. The excess in the SMG composite, when fitted by M82, is consistent with an additional power-law component emission from an AGN (dotted line; see \S\ref{agn}). 
}
\end{figure*}

\section{Observations and Reduction}

We observed each target using the low resolution Long-Low (LL) observing mode of IRS ($R \sim 57-126$) at two different nod positions for 30 cycles of 120s each. We cover restframe emission longwards of 6\,$\mu$m to probe for PAH emission at 6.2, 7.7, 8.6 and 11.3\,$\mu$m and for silicate absorption centered at 9.7\,$\mu$m. The data were obtained between December 2005 and March 2006. 

The data were processed using the Spitzer IRS S13 pipeline\footnote{http://ssc.spitzer.caltech.edu/irs/dh/}, which includes saturation flagging, dark subtraction, linearity correction, ramp correction and flat-fielding. With a slit size of $\sim 10.5 \times 168 \arcsec$, IRS does not resolve the SMGs spatially, and the targets were treated as point sources throughout the data reduction and analysis.  We performed additional reduction of the 2D spectra using IRSCLEAN\,\footnote{http://ssc.spitzer.caltech.edu/archanaly/contributed/irsclean} to remove rogue pixels, and relied on differencing between the nod positions to subtract the residual background. We used the SPitzer IRS Custom Extraction (SPICE)\,\footnote{http://ssc.spitzer.caltech.edu/postbcd/spice.html} software to optimally extract flux-calibrated 1D spectra, by taking a weighted average of profile-normalized flux at each wavelength to increase the S/N of these faint sources. 

\section{Results and Discussion}\label{qualitative}

The mid-IR spectra of SMM\,J221733, SMM\,J163659, SMM\,J030228 and SMM\,J163639 show moderate to strong PAH features (see Fig.\,1), and we refer to these targets collectively as the \textit{PAH sample}.  Detection of PAH emission is assumed to indicate the presence of SB activity. At most a very shallow dip is present around 9.7\,$\mu$m in the spectra, indicating little silicate absorption. 

Our highest-$z$ source, SMM\,J163650, is somewhat different to the other targets, with a broad feature at restframe $\sim$8\,$\mu$m, unlike the typical blended PAH complex of the 7.7- and 8.6-$\mu$m features found in SB galaxies.  It is more reminiscent of the spectrum of Mrk\,231 (\citealt{armus06b}, A06b), which features an unabsorbed continuum between absorption from silicates at longer wavelengths and hydrocarbons at shorter ones (\citealt{spoon04}, S04; \citealt{weedman06}). This similarity suggests that SMM\,J163650 has more substantial AGN-activity than the SMGs in the PAH sample, as expected from the presence of a strong CIV ($\lambda 1549$) feature at restframe UV (C05) and a broad H$\alpha$ component ($\simeq 1753 \pm 238$ km s$^{-1}$; \citealt{swinbank04}), both revealing the unambiguous presence of an AGN.  We discuss the properties of this source in more detail in a subsequent paper discussing the full sample (Men\'{e}ndez-Delmestre et al. in prep.) and concentrate here on the median properties of the SMGs with clear PAH emission.

To get an insight into the physics inherent to SMGs in our PAH sample, we compare their spectra with extensively studied local templates: the AGNs Mrk\,231 (A06b) and NGC\,1068 \citep{sturm00}, the SB M82 \citep{forster-schreiber03} and the ULIRGs Arp\,220 (A06b) and NGC\,6240 (\citealt{armus06a}, A06a). Arp\,220 has been a favorite template for high-redshift SMGs (\citealt{pope06, kovacs06}): it has strong PAH features, indicative of SB activity, and a steep mid-IR continuum due to a heavily obscured nuclear component inferred to be responsible for the bulk of the IR luminosity (S04). AGNs have been identified in both merging components of NGC\,6240 but SB dominates the total IR luminosity \citep{komossa03}.  

A qualitative comparison of the spectra of our PAH sample with these templates rules out Mrk\,231, NGC\,1068 and Arp\,220 as good matches, but the spectra are similar to those of M82 and NGC\,6240. Similar results were found by L05, who detected strong PAH features in the spectra of two luminous SMGs at $z \sim 2.8$ that were well fitted by an M82-type spectrum.

\begin{figure}\label{ratios}
\plotone{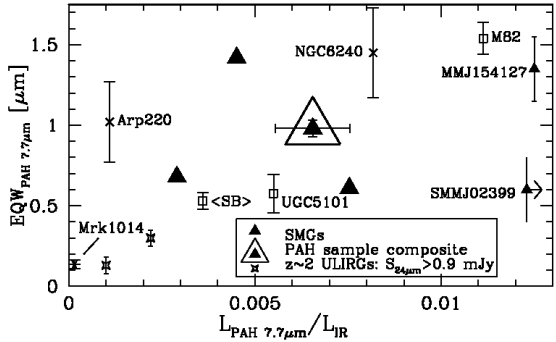}
\caption{Relative strengths of the 7.7-$\mu$m PAH feature as measured by the PAH-to-IR ($8-1000\,\mu$m) luminosity ratio and the restframe EW for the SMGs in the PAH sample (large triangles). We estimate the values for M82 \citep{forster-schreiber03}, Arp\,220 (S0), NGC\,6240 (A06a) and for the L05 pair of SMGs (small triangles) directly from their published spectra assuming a linear continuum slope; the error bars reflect the uncertainties of this approach. We derive a lower EW for NGC\,6240 than A06a, due to differences in continuum definition. Error bars for the Y05 sources, for the SB-dominated UGC5101 \citep{armus04} and for the average of 13 nearby SB galaxies studied with IRS \citep{brandl06} reflect the uncertainties presented by the authors of the respective papers.}
\end{figure}

\subsection{The Composite SMG spectrum}\label{analysis}

We take advantage of the similarity between the spectra in the PAH sample and of our precisely known redshifts (C05) to double our S/N constructing a composite spectrum by averaging the individual spectra (Fig.\,1). We use the composite spectrum to make a preliminary assessment of the independent contributions of SB and AGN activity in our PAH sample. Normalizing the local templates to the 7.7-$\mu$m peak in the composite spectrum, we find that the composite spectrum is well fitted at $\lambda \lesssim 9 \mu$m by the NGC\,6240 and M82 spectra (Fig.\,1). However, neither template provides a good fit to the composite spectrum at $\lambda \gtrsim 9 \mu$m:  the continuum emission of NGC\,6240 exceeds that of the composite spectrum at $\lambda \gtrsim 12 \mu$m, while the spectrum of M82 falls below it. No physically reasonable additional AGN component can be added to the NGC\,6240 spectrum to produce a good fit  to the composite spectrum at longer wavelengths. On the other hand, an M82-type spectrum plus a power-law continuum provides a good fit to the composite SMG data at all wavelengths.

\subsubsection{Starburst Component}

The 7.7-$\mu$m PAH feature is generally the most prominent in the mid-IR spectra of SB galaxies. Its strength relative to the continuum, measured by the equivalent width (EW), can be used to evaluate the fractional SB contribution to the total bolometric output, as the hot mid-IR continuum is enhanced significantly in the presence of an AGN. SB-dominated objects, such as M82 and NGC\,6240, are characterized by larger PAH EWs than objects with a prominent AGN, such as Mrk\,231.  

EWs are sensitive to how the continuum is defined.  We define a linear continuum by interpolating  between two points clear of PAH emission, at 6.8\,$\mu$m and 13.7$\mu$m, or at 9\,$\mu$m when the spectrum does not include one of these points. In  Fig.\,2 we plot the 7.7-$\mu$m restframe EWs and PAH-to-IR luminosity ratios for the SMGs in the PAH sample with 7.7-$\mu$m coverage and for the composite spectrum.  The error in $L_{7.7\mu\small\textrm{m}}$/$L_{\small\textrm{IR}}$ for our SMG sample is dominated by a $\simeq 20\%$ uncertainty in the IR luminosities (C05)\footnote{\citet{kovacs06} show that SMGs fall below the local FIR-radio relation and thus the C05-derived $L_{\small\textrm{IR}}$ values, which rely on this relation, are in average overestimated by a factor of $\sim 2$.}. We compare a number of low- and high-$z$ sources, including 2 ULIRGs at $z \sim 2$ with clear PAH detections from the Y05 sample with S$_{24\mu\small\textrm{m}} \gtrsim 0.9$ mJy, and 2 SMGs (SMM\,J02399$-$0136 with S$_{850\mu\small\textrm{m}} =$\,23\,mJy and MM\,J154127+6616 with S$_{850\mu\small\textrm{m}} =$\,14.6\,mJy) at $z \simeq 2.8$ (L05). 

According to the line-to-continuum (l/c) diagnostic presented by \citet{genzel98}, systems with (l/c)$_{7.7\mu m}$\,$\gtrsim$ 1 are classified as SB$-$dominated and those with (l/c)$_{7.7\mu m}$\,$\textless$ 1, as AGN-dominated. With (l/c)$_{7.7\mu\small\textrm{m}}$\,$\gtrsim$ 1, SBs appear to dominate the Y05, L05 and our PAH sample\footnote{For our sample, $\textrm{(l/c)} \sim 1$ corresponds to $\textrm{EQW} \sim 0.5\,\mu$m}. However, the distribution in 7.7-$\mu$m EW and PAH-to-IR luminosity ratio in Fig.\,2 may suggest a distinction in the relative SB-to-AGN contributions, with lower values of these parameters indicating a stronger AGN contribution. We distinguish three regions in  Fig.\,2: (1) a region with low PAH-to-IR luminosity ratios, occupied by Mrk\,1014 \citep{armus04} and the 24\,$\mu$m-bright sample of Y05; (2) an intermediate PAH-to-IR luminosity region where NGC\,6240 and the bulk of the SMGs in our sample are located; and (3) a region with the highest PAH-to-IR luminosity ratios, occupied by M82 and the two SMGs in L05. 

At $z\sim2$, 24-$\mu$m flux traces 8-$\mu$m restframe continuum; a stronger hot mid-IR continuum (produced by an AGN) dilutes the strength of PAH features, leading to lower $L_{7.7\mu\small\textrm{m}}/L_{\small\textrm{IR}}$. The location of the Y05 sample in the plot could follow from the selection of 24\,$\mu$m-bright targets (S$_{24\mu m} \gtrsim$ 0.9 mJy) at this redshift, which would select objects with lower SB-to-AGN ratios. SMGs in our sample have higher $L_{7.7\mu\small\textrm{m}}/L_{\small\textrm{IR}}$ ratios, similar to NGC\,6240, which we interpret as an indication of a markedly stronger SB contribution to the total luminosity than the Y05 sample. With similar IR-luminosities, the location of the L05 SMG pair in this plot indicates that the 7.7-$\mu$m PAH feature is very strong. With large values for both the EW and the PAH-to-IR luminosity ratio, MM\,J154127 has been suggested to be dominated by SB-activity (L05). The large PAH-to-IR luminosity ratio for SMM\,J02399 suggests strong SB-activity; however, the relatively low EW value, together with the evident strong mid-IR continuum (see Fig.\,1, L05) is consistent with this source having roughly equal AGN and SB contributions.  

The SMGs in our PAH sample have values of EW$_{7.7\mu\small\textrm{m}}$ and $L_{7.7\mu\small\textrm{m}}/L_{\small\textrm{IR}}$ that place their SB-to-AGN ratio between that of the AGN-dominated ULIRG Mrk\,1014 and the SB M82. This is qualitatively similar to NGC\,6240, which has both SB and AGN components. As a caveat, we note that even though the 7.7-$\mu$m PAH-to-IR luminosity ratio is associated to the SB-to-AGN ratio, it is also sensitive to details of the spectral energy distribution of the system, such as the presence of multiple dust components at different temperatures and the amount of extinction (e.g. Arp\,220; S04). This may explain the particularly high 7.7-$\mu$m PAH-to-IR luminosity ratio for SMM\,J02399.

\subsubsection{The AGN Component}\label{agn}

Mid-IR line diagnostics suggest that SMGs are SB-like.  The spectra of the SB-dominated galaxies M82 and NGC\,6240 provide a good fit to the composite spectrum at $\lambda \lesssim 9 \mu$m; however, only an M82-type spectrum with an additional power law AGN component gives a good fit to the composite spectrum at all wavelengths. The power-law component is defined as $S_\nu \sim \nu^{-2.9}$, consistent with the range of IR spectral indices for 3C quasars in \citet{simpson00}. 

From the power-law component flux at 10.5\,$\mu$m we estimate the X-ray luminosity ($L_{X}$) using the correlation between $S_{10.5\mu \small\textrm{m}}$ and $S_{2-10 \small\textrm{keV}}$ presented by \citet{krabbe01}.  This yields $L_{X}\sim10^{44}$\,erg s$^{-1}$ for an AGN at the average redshift for the SMGs presented in this paper, z $\sim$ 1.4, in reasonable agreement with the X-ray luminosities found for the SMGs in A05. Following A05's approach, we compare the average X-ray-to-far-IR ratio of the SMGs in the PAH sample with the typical ratio for quasars and find that the residual flux is consistent with an underlying AGN contributing in the order $\sim$10$\%$ to the total far-IR emission.  This agrees with the A05 result that AGN activity is often present in SMGs but does not dominate the energetics.  Since the 10.5-$\mu$m excess is dominated by lower redshift sources, further SMGs at $z \lesssim 2$, included in Men\'{e}ndez-Delmestre et al. (in prep), will better constrain this excess.

\section{CONCLUSIONS}

We present first results of a \textit{Spitzer} program to characterize the mid-IR spectra of high redshift SMGs. We compare the spectra to well-studied local templates and find that SMGs have starburst mid-IR spectra more like M82 than the often quoted local analog Arp\,220. The composite spectrum of the SMGs in the PAH sample is well fitted by an M82-like starburst-component with a power-law continuum most likely representing a fainter underlying AGN.  This similarity to the M82 spectrum suggests that the chemistry of the interstellar medium and radiation fields in these systems may be understood by looking at local galaxies in detail. Analysis of the 7.7-\,$\mu$m equivalent widths and PAH-to-IR luminosity ratios show that SMGs are markedly different from 24\,$\mu$m-selected samples, such as the $z\sim2$ sample of Y05, which show stronger AGN contributions. This work provides further evidence that SMGs host both star-formation and AGN activity, but that star-formation dominates the bolometric luminosity, reiterating the role of SMGs as the build-up sites for a significant fraction of the stellar content we see today. 

By probing the lower-redshift end of the C05 SMG sample distribution, this sample provides rest-frame wavelength coverage longwards of $9\,\mu$m to assess the AGN contribution. The SMG at $z\,=\,2.38$, with a redshift closer to that of a typical C05 SMG, displays a potentially more AGN-dominated Mrk\,231-type broad feature at restframe $\sim$8\,$\mu$m. The difference in AGN contributions within our preliminary sample suggests an increasing relative AGN-activity in SMGs at higher redshifts, potentially due to the $24 \mu m$ flux limit applied to the sample. The full sample, with a more extended redshift distribution, will provide us with additional valuable information concerning the typical SMG population.

\acknowledgements
We thank our anonymous referee for valuable comments. We thank the Spitzer Science Center staff for their support, particularly Patrick Ogle for his help in the optimization of the spectral extraction.  AWB thanks the Research Corporation and the Alfred P. Sloan Foundation. DMA and IRS acknowledge support from the Royal Society.  This work is based on observations made with the Spitzer Space Telescope, which is operated by the Jet Propulsion Laboratory, California Institute of Technology under a contract with NASA. Support for this work was provided by NASA through an award issued by JPL/Caltech.

\end{document}